\documentstyle[aps,prb]{revtex}
\begin{document}
\tighten
\title{Quantum effects in a  superconducting glass model}
\author{T.K. Kope\'{c}$^\ast$ and J.V.Jos\'{e}}
\address{ Department of Physics, Northeastern University,
Boston, Massachusetts 02115}
\date{\today}

\maketitle
\begin{abstract}
We study disordered Josephson junctions arrays
with long-range interaction and  charging effects. The model  consists
of two orthogonal sets of  positionally disordered $N$ parallel
filaments (or wires) Josephson coupled at each crossing
and in the presence of a homogeneous and transverse magnetic field.
The large charging energy (resulting from small self-capacitance of the
ultrathin wires) introduces important quantum fluctuations of
the superconducting phase within each filament.
Positional disorder and magnetic field frustration
induce spin-glass like ground state, characterized by not having
long-range order of the phases.
The stability of this phase is destroyed for sufficiently large
charging energy. We have evaluated the temperature vs charging energy
phase diagram by extending the methods developed
in the theory of infinite-range spin glasses,
in the limit of large magnetic field.
The phase diagram in the different temperature regimes
is evaluated by using variety of methods, to wit:
 semiclassical WKB and variational methods,
Rayleigh-Schr\"{o}dinger  perturbation theory and pseudospin
effective Hamiltonians. Possible experimental consequences
of these results are briefly discussed.
\end{abstract}
%%%%%%%%%%%%%%%%%%%%%%%%%%%%%%%%%%%%%%%%%%%%%%%%%
\narrowtext
%%%%%%%%%%%%%%%%%%%%%%%%%%%%%%%%%%%%%%%%%%%%%
\section{Introduction and model Hamiltonian}
%%%%%%%%%%%%%%%%%%%%%%%%%%%%%%%%%%%%%%%%%%%%%%
Josephson junction arrays (JJA) have been the subject
of considerable recent
research activity\cite{expov}. Such systems  consist
of superconducting  grains embedded in a nonsuperconducting host
and coupled together by the Josephson effect.
Recently there has been a surge of both theoretical and experimental
interest in studying
 disordered Josephson-coupled systems in an applied magnetic
field\cite{recent,recent2,recent3,joseglass}.
Theoretically, it has been shown that in the presence
of sufficiently strong magnetic fields the system may freeze into a state
exhibiting spin-glass-like type order among the superconducting grains
\cite{theory}. From the experimental viewpoint the interest is
motivated by the existence of irreversibility lines in the
temperature-field diagram, in
virtually all high-$T_c$ superconductors thus  suggesting
the existence of a glassy phase\cite{experim}.
\par

For sufficiently small  grains, the behavior of each junction
in the JJA modified by the quantum effects which  arise from the
small capacitances that lead to large charging energies.
  The competition between  phase coherent ordering and charging effects
in periodic JJA has been the  subject of a number of studies\cite{and}
 and it is  by now well established that
 for sufficiently large charging energy  quantum phase fluctuations
lead to the complete suppression of long-range  superconducting
order\cite{sim4}.
\par

In disordered JJA with small capacitances
thermal, random  {\it and} quantum fluctuation will
determine  the physics of the system and an interesting question arises
regarding the competition between them.  The problem
we would like to address in this paper is then:
What is the effect of having a competition between the thermal, quantum
and random fluctuations on the long range properties of a
Josephson junction network. This is a highly nontrivial question in
general but partial answers can be obtained in certain limits.
Since there are  only very limited studies on this issue,
(see, eg Ref. \onlinecite{choi}),  the purpose of this paper
is to investigate these quantum-fluctuation effects systematically
in a mean-field like approximation, to be specified below.
We shall study a quantum  model of a  disordered JJA
system which, in principle, could be realized experimentally
and simultaneously allowing for a detailed theoretical treatment
thus constituting an attractive setting for the study of the
complicated interplay between quenched disorder, interactions
and quantum fluctuations.
\par
To be  specific, we will study  a stack of two sets of $N$ mutually
perpendicular parallel wires (or filaments)  Josephson coupled  at
nodes (see, Fig.\ref{fig1}). Since  each horizontal (vertical) filament
of the system is directly coupled to every other vertical (horizontal)
filament,  the number of nearest neighbors $z$ in this model is $z=N$.
This is then a realization of a JJA with {\it long-range}
(infinite-range, in the thermodynamic limit $N\to\infty$)
interactions which differ from the conventional 2D Josephson
junctions arrays. Furthermore, we assume that the distance between
neighboring parallel wires varies randomly around some average
value $l$. Finally, the system is placed in a transverse magnetic field
$B$ and we shall assume that the Josephson couplings are sufficiently
small so that the induced magnetic fields are negligible in
comparison with $B$ so that the phase gradient along any
filament results only from the presence of the external  magnetic field.
The dynamical  variables of this system are the
superconducting phases associated with each wire.
The properties of the  (classical) model defined above
 have been studied some time ago\cite{recent}.
Very recently the thermodynamical properties of this system
(both periodic and positionally disordered version)
have been investigated  theoretically and experimentally
\cite{wires1,wires2,wires3}.

\par
An important ingredient in our present considerations is that
we allow for {\it quantum} phase fluctuation within  a wire
assuming the node junctions have a sufficiently small
self-capacitance $C$. In this case
the charging energy becomes a dominant quantity
considerably affecting the properties of the array  in the
low temperature regime. More precisely, we are interested in the
behavior where the quantum fluctuations compete with the
formation of the superconducting glassy phase due to
randomness and magnetic frustration.
\par
In the low temperature regime we expect to have
a glassy phase with randomly frozen superconducting phases.
To examine the extent to which the system is frozen, it is
convenient to introduce the Edwards-Anderson order
parameter\cite{Edwards} defined by
\begin{equation}
q_{EA}\equiv [\langle {\bf S}_i\rangle_T^2]_{\rm av},
\end{equation}
where
\begin{equation}
{\bf S}_i=[S^x_i,S^y_i]\equiv [\cos(\phi_i),\sin(\phi_i)],
\label{spin}
\end{equation}
while $\langle\dots\rangle_T$ and
$[\dots]_{\rm av}$ denote thermodynamic
and configurational averaging, respectively.
In a  disordered state  the phases will randomly sample
their entire phase space and $q=0$, while in a
frozen system $q\neq0$ and $[\langle {\bf S}_i\rangle_T]_{\rm av}=0$
indicating the absence of long-range order.
The spin glass-like  ground state may  be destabilized
by quantum fluctuations\cite{quantum}. As one varies the strength
of the charging energy  (i.e., increasing the
strength of the quantum fluctuations) there can be a phase transition at
zero temperature between the  glassy  phase and the
paracoherent disordered ground states.
This {\it quantum phase transition} in random spin systems
has received much attention recently\cite{quantum2}.
However, its  corresponding nature
in quantum disordered Josephson-coupled systems
has not been previously studied. The goal of this paper is
to examine this problem in some detail starting from a
quantum gauge--glass model defined in Eq. (\ref{hamil}) below.
\par
As discussed in the main body of the paper, we need to use different
complementary calculational techniques to attack the problem, for
each one of them is by nature approximate and their regions of
validity are different. It is also important that using only one of these
techniques by itself may lead to spurious results that can only be
validated by an independent check. A case in point will be the
reentrant transition found
in the variational calculation, which is valid in the semiclassical
region, and which is not found in our low temperature expansion.
\par
We now describe the main   body of the paper.
In Section II we define the model Hamiltonian, we perform
the quenched average over the disorder,
followed by a functional integral formulation of the
mean--field theory and the derivation of the saddle--point equations
in the $N\to\infty$ limit. This  constitutes an $\underline{exact}$
self-consistency condition for the  glass order parameters.
Section III presents the phase diagram for the model
preceeded by an exhaustive investigation of the self-consistency
equations via a variety of approaches including:
semiclassical WKB and variational methods,
Rayleigh-Schr\"{o}dinger perturbation theory and
pseudospin effective Hamiltonians in a truncated charge Hilbert space.
Finally, in Section IV, we discuss  our results and present our
conclusions.

%%%%%%%%%%%%%%%%%%%%%%%%%%%%%%%%%%%%%%%%%%%%%%%%%%%%%%%%%%%
\section{The model}
%%%%%%%%%%%%%%%%%%%%%%%%%%%%%%%%%%%%%%%%%%%%%%%%%%%%%%%%%%
The {\it quantum}  Hamiltonian of the disordered system of
Josephson-coupled wires
is given by
\begin{eqnarray}
H&&=H_C+H_J,
\nonumber\\
H_C&&=-\frac{K}{2}\sum_{i=1}^N\left(\frac{\partial^2}{\partial\phi_{hi}^2}
+\frac{\partial^2}{\partial\phi_{vi}^2}\right),
\nonumber\\
H_J&&=\frac{E_J}{\sqrt{N}}\sum_{i=1}^N
\sum_{j=1}^N \left[1-\cos(\phi_{hi}-\phi_{vj}-f_{ij})\right].
\label{hamil}
\end{eqnarray}
\narrowtext
Here, $\hat{n}_i=({2e}/{i}){\partial}/{\partial\phi_i}$ is the
charge operator while
$\phi_{hi}$ ($\phi_{vj}$) represent the superconducting
phase of the $i-$th horizontal ($j-$th vertical) wire, respectively;
$E_J$ is the Josephson coupling  and  $K=4e^2/C$ the charging,
respectively.
In order to have a well defined thermodynamic limit we have to scale
the Josephson coupling energy by a factor $\sqrt{N}$.
Furthermore, $f_{ij}=({2\pi}/{\Phi_0})
\int_{{\bf r}_i}^{{\bf r}_j}{\bf A}\cdot d{\bf l}$
is the line integral of the vector potential ${\bf A}$
and ${\Phi_0}$ is the elementary flux quantum.
\par
It is clear that   $E_J$ is  a positive quantity
 (and it  may depend only
on the distance between wires which we are neglecting)
and  it does not introduce
any frustration in the system.  Magnetic field
{\it and} random location of the wires is what
 generates variations
of the phase  parameters $f_{ij}$, thus allowing
for the random frustration present in the system\cite{remark1}.
The relevant quantity is thus the effective random coupling
matrix in the Josephson part of the Hamiltonian (\ref{hamil})
\begin{equation}
{\bf J}_{ij}=\frac{E_J}{\sqrt{N}}
\left(
\begin{array}{cc}
 0 & e^{if_{ij}} \\
 e^{-if_{ji}} & 0 \\
\end{array}
\right),
\label{ranmat}
\end{equation}
whose density of the eigenvalues is given by
\begin{equation}
\rho(E)=-({1}/{\pi})
\mbox{Im}\overline{\left[{E\delta_{ij}-{\bf J}_{ij}
+i0}\right]_{ii}^{-1}},
\label{density}
\end{equation}
where the bar denotes averaging over the positional disorder.
The behavior of (\ref{density}) varies with increasing the
strength of the magnetic field.
In the case of large magnetic field, so that
the flux  per average plaquette $\Phi =Bl^2$ is much larger
then the elementary flux quantum $\Phi/\Phi_0 \gg 1$,the
frustration parameters $f_{ij}$ acquire random values
and fill the interval $(0,2\pi]$ uniformly.
In this limit the effects of disorder become
especially apparent and the behavior moves toward the
asymptotic regime which is essentially field independent\cite{recent}.
In this case  correlations between the matrix elements
${\bf J}_{ij}$ vanish and
the density of states (\ref{density}) approaches the
Wigner semicircular law\cite{recent}
\begin{equation}
\rho(E)\to\rho_W(E)=
{(\pi E_J)}^{-1}\sqrt{1-\left({E}/{2E_J}\right)^2},
\end{equation}
implying that the random matrix (\ref{ranmat})
belongs to the Gaussian unitary ensemble with
only the second moment non-vanishing,
\begin{equation}
\langle {\bf J}_{ij}{\bf J}_{kl}\rangle_{\rm av}
=\frac{E_J}{N}\delta_{ik}\delta_{jl}.
\label{average}
\end{equation}
Therefore, in the high-field regime,
we can implement a mean-field theory
of our quantum gauge-glass  problem in a way closely resembling
the infinite-range interaction Sherrington-Kirkpatrick
magnetic spin-glass model\cite{remark2}. However, the resulting
formulation is  not  just a quantum extension of  the
planar spin  glass model. Unlike the random bond XY
spin glass, the improper global rotation $\phi_i\to-\phi_i$
(i.e. time reversal) is {\it not} a symmetry of our quantum
gauge-glass model. To make this observation apparent we introduce the
two component ``spin" vector (\ref{spin})
 so that the Josephson part of the Hamiltonian (\ref{hamil})
reads
\widetext
\begin{eqnarray}
H_J=
-\frac{E_J}{2\sqrt{N}}\sum_{i=1}^N\sum_{j=1}^N
\left[e^{if_{ij}}\left({\bf S}_{hi}
\cdot{\bf S}_{vj}-i{\bf\hat{ z}}\cdot{\bf S}_{hi}\times{\bf S}_{vj}\right)
+{\rm h.c.}\right],
\label{dm}
\end{eqnarray}
\narrowtext
where ${\bf\hat{ z}}$ is a unit vector perpendicular to the plane
containing ${\bf S}_i$. We note that, in addition to the conventional
XY coupling ${\bf S}_{hi}\cdot{\bf S}_{vj}$ there is a cross term
${\bf S}_{hi}\times{\bf S}_{vj}$ which is the analogue of the
spin-orbit (i.e. Dzialoshinsky-Moriya (DM)) interaction in magnetic
systems --  essentially violating the time-reversal symmetry\cite{gingras}.
Thus, the present quantum gauge--glass problem formally resembles
more closely the quantum spin--glass formulation in the presence
of the DM anisotropy\cite{kopec1}.

%%%%%%%%%%%%%%%%%%%%%%%%%%%%%%%%%%%%%%%%%%%%%%%%%%%%%%%%%%%%
\section{Disorder average  and mean-field
formulation}
%%%%%%%%%%%%%%%%%%%%%%%%%%%%%%%%%%%%%%%%%%%%%%%%%%%%%%%%%%%%
 In a random system we need
to calculate the average of the free energy density
${\cal F}=-(1/\beta)\ln Z/2N$ over the disorder
(\ref{average}). This is done by  using the replica method permitting to
average the replicated partition function $Z^n$ instead of
$\ln Z$,
\begin{equation}
\beta {\cal F} =-\lim_{N\rightarrow\infty}\lim_{n\rightarrow 0}
\frac{1}{nN}(\langle{Z^n}\rangle_{\rm av} -1).
\label{fen}
\end{equation}
%%%%%%%%%%%%%%%%%%%%%%
It is convenient to express the replicated
 partition function $Z^n=Tr\exp(-\sum_\alpha\beta H^\alpha)$
in the interaction representation as
\begin{eqnarray}
Z^n={\rm Tr}e^{-\beta\sum_\alpha H_0^\alpha}
T_\tau\exp\left[- \sum_\alpha\int_{0
}^{\beta}d\tau{H}_I^\alpha(\tau)  \right],
\label{exp}
\end{eqnarray}
%%%%%%%%%%%%%%%%%%%%%%
with the interaction picture Hamiltonian
\begin{equation}
{H}_I^\alpha(\tau) =e^{\tau H_0^\alpha}
{H}_I^\alpha e^{-\tau H_0^\alpha},
\end{equation}
and the free part
\begin{eqnarray}
H_0^\alpha&&=-\frac{K}{2}\sum_{i}\left(
\frac{\partial^2}{\partial\phi_{hi}^{\alpha 2}}+
\frac{\partial^2}{\partial\phi_{vi}^{\alpha 2}}
\right),
\end{eqnarray}
where $H^\alpha=H_0^\alpha+{H}_I^\alpha$ is the total
Hamiltonian. For the interaction Hamiltonian
one has explicitly
\begin{eqnarray}
H_I^\alpha(\tau)&&= \frac{E_J}{2\sqrt{N}}\left\{
\sum_{i=1}^N\sum_{j=1}^N e^{if_{ij}}\left[{\bf S}_{hi}^\alpha(\tau)\cdot
{\bf S}_{vj}^\alpha(\tau) -i{\bf z}
\cdot{\bf S}_{hi}^\alpha(\tau)\times{\bf
S}_{vj}^\alpha(\tau)\right] +{\rm h.c}\right\},
\end{eqnarray}
\narrowtext
so that the statistical average can be taken in the ensemble given
by $H_0^\alpha$. Here $T_\tau$ is the Matsubara ``imaginary time" ordering
operator allowing us to treat the time dependent operators
${\bf S}_i^\alpha(\tau)=e^{-\tau{H}_0}
{\bf S}_i^\alpha e^{\tau{H}_0}$ as $c$-numbers within the
time-ordered exponential (\ref{exp}). Consequently, the
Gaussian average (\ref{average}) readily gives
%%%%%%%%%%%%%%%%
\begin{equation}
{\langle Z^n\rangle}_{\rm av} =  {\rm Tr}e^{-\sum_\alpha\beta H_0^\alpha}
T_\tau\exp\left[-\int_{0}^{\beta}d\tau\int_{o}^{\beta}d\tau'
{\Omega}(\tau,\tau') \right]
\label{zn}
\end{equation}
where
\widetext
%%%%%%%%%%%%%%%%%%%%%%%%%
\begin{eqnarray}
{\Omega}(\tau,\tau')& =& \frac{E_J^2}{4N}
\sum_{i=1}^N\sum_{j=1}^N\sum_{\alpha\beta}
\{[{\bf S}_{hi}^\alpha(\tau)
\cdot{\bf S}_{vj}^\alpha(\tau)][{\bf S}_{hi}^\beta(\tau')\cdot
{\bf S}_{vj}^\beta(\tau') +\nonumber\\
& & + [{\bf S}_{hi}^\alpha(\tau)
\times{\bf S}_{vj}^\alpha(\tau)]\cdot
[{\bf S}_{hi}^\beta(\tau')\times{\bf S}_{vj}^\beta(\tau')]\}.
\label{omega}
\end{eqnarray}
\narrowtext
To make further progress we utilize the vector identity
$({\bf A}\times{\bf B})\cdot ({\bf C}\times{\bf D}) =
({\bf A}\cdot{\bf C})({\bf B}\cdot{\bf D}) - ({\bf B}\cdot{\bf C})
({\bf A}\cdot{\bf D})$ to reduce mixed vector products
along with integrations over auxiliary variables
according to the formulas
%%%%%%%%%%%%%%%%%%%%%%%%%%%%%
\begin{eqnarray}
\exp(\pm ab)&&=\int\frac{dxdy}{2\pi i}\exp[\pm (xy-ax-by)],
\nonumber\\
\exp(a^2/2)&&=\int\frac{dx}{\sqrt{2\pi}}\exp(-x^2/2-ax),
\end{eqnarray}
%%%%%%%%%%%%%%%%%%
allowing to decouple various quartic interactions in Eq.(\ref{omega})
 and reduce
Eq.(\ref{zn}) to the effective single-filament problem.
In terms of the functional integrals
\widetext
\begin{equation}
\langle Z\rangle_{\rm av} = \int
\prod_{\alpha\beta}
\prod_{\mu\nu}DQ^{\alpha\beta}_{\mu\nu}DR^{\alpha\beta}
DP^{\alpha\beta}_{1\mu\nu}
DP^{\alpha\beta}_{2\mu\nu} \exp(-2NL[{\bf P}_1,{\bf P}_2,{\bf  Q,R}] ),
\end{equation}
\narrowtext
involving the non-local (in time) tensor fields
$X^{\alpha\beta}_{\mu\nu}(\tau,\tau')$ $(\bf X\equiv P_1,P_2,Q)$ and
$R^{\alpha\beta}(\tau,\tau')$ where $\mu,\nu=x,y$.
The   effective local Lagrangian reads
\widetext
\begin{equation}
L[{\bf P}_1,{\bf P}_2,{\bf  Q,R}]
 = \mbox{Tr}{\bf P }_1{\bf P }_2+ \mbox{Tr}{\bf Q}^2 + \mbox{Tr}{\bf R}^2
 - \ln\Phi[{\bf P}_1,{\bf P}_2,{\bf  Q,R}],
\end{equation}
\narrowtext
with
\begin{eqnarray}
\mbox{Tr}{\bf X}{\bf Y}&&= \int_{0}^{\beta}d\tau
\int_{0}^{\beta}d\tau' \sum_{\alpha\beta} \sum_{\mu\nu}
X^{\alpha\beta}_{\mu\nu}(\tau,\tau')Y^{\beta\alpha}_{\nu\mu}(\tau',\tau),
\nonumber\\
\mbox{Tr}{\bf R}^2&&= \int_{0}^{\beta}d\tau
\int_{0}^{\beta}d\tau' \sum_{\alpha\beta}
R^{\alpha\beta}(\tau,\tau')R^{\beta\alpha}(\tau',\tau).
\end{eqnarray}
Here,
\widetext
\begin{equation}
\Phi[{\bf P}_1,{\bf P}_2,{\bf  Q,R}]
=  {\rm Tr}e^{-\beta H_0}
T_\tau\exp\left[-\int_{0}^{\beta}d\tau\int_{0}^{\beta}d\tau'
\hat{H}_{\rm eff}(\tau,\tau') \right],
\end{equation}
\narrowtext
is  the effective time-dependent
single-filament Hamiltonian describing interactions
between replicas
\begin{eqnarray}
&&\hat{H}_{\rm eff}(\tau,\tau')  = -E_J \sum_{\alpha\beta}
\sum_{\mu\nu}\left[Q^{\alpha\beta}_{\mu\nu}(\tau,\tau')
-\frac{1}{2}P^{\alpha\beta}_{1\mu\nu}(\tau,\tau')+\right.\nonumber\\
&&\left.R^{\alpha\beta}(\tau,\tau')\delta_{\mu\nu}
-\frac{1}{2}P^{\alpha\beta}_{2\nu\mu}(\tau,\tau')
\right]S^\alpha_{\mu}(\tau)S^\beta_{\nu}(\tau').
\end{eqnarray}
In the thermodynamic limit, $N \rightarrow \infty$,
the steepest descent method
can be used which amounts to finding the stationary points
$X^{\alpha\beta}_{0,\mu\nu}$ and $R^{\alpha\beta}_0$ determined by the
extremal conditions
\begin{eqnarray}
\delta L[{\bf P}_1,{\bf P}_2,{\bf  Q,R}]/
\delta X^{\alpha\beta}_{\mu\nu} &=& 0,\nonumber\\
\delta L[{\bf P}_1,{\bf P}_2,{\bf  Q,R}]/\delta R^{\alpha\beta} &=& 0.
\end{eqnarray}
Thus, one obtains
\begin{eqnarray}
Q^{\alpha\beta}_{0,\mu\nu}(\tau-\tau') &&= \frac{E_J}{2}
G^{\alpha\beta}_{\mu\nu}(\tau-\tau'),\nonumber\\
R^{\alpha\beta}_0(\tau-\tau') &&= \frac{E_J}{2}
 \sum_\mu G^{\alpha\beta}_{\mu \mu}(\tau-\tau'),\nonumber\\
P^{\alpha\beta}_{0,1\mu\nu}(\tau-\tau')&&=
P^{\alpha\beta}_{0,2\nu\mu}(\tau-\tau')
=\frac{E_J}{2}
G^{\alpha\beta}_{\mu\nu}(\tau-\tau')
\end{eqnarray}
where the correlation function   describing the
dynamic  self-interaction is
\widetext
\begin{equation}
G^{\alpha\beta}_{\mu\nu}(\tau-\tau') =
 \frac{\displaystyle{\rm Tr}e^{-\beta H_0}
T_\tau S^\alpha_{\mu}(\tau)S^\beta_{\nu}(\tau')
\exp\left[-\int_{0}^{\beta}d\tau_1\int_{0}^{\beta}d\tau_2
\hat{H}_{\rm eff}(\tau_1,\tau_2) \right] }
{\displaystyle{\rm Tr}e^{-\beta H_0}
T_\tau\exp\left[-\int_{0}^{\beta}d\tau_1\int_{0}^{\beta}d\tau_2
\hat{H}_{\rm eff}(\tau_1,\tau_2) \right]}.
\label{matrix}
\end{equation}
\narrowtext
Equation (\ref{matrix}) represents an $\underline{exact}$
self--consistency condition
for the replica dependent matrix Green function
$G^{\alpha\beta}_{\mu\nu}(\tau-\tau')$ of the quantum disordered
Josephson model (\ref{hamil}).
For classical spin glasses,
 this is a matrix $G^{\alpha\beta}$,
where the off--diagonal components of $q_{\alpha\beta}$ can be related to the
spin--glass--like Edwards Anderson order parameter
\begin{eqnarray}
q_{EA}&&={\rm max}_{\alpha\neq\beta}(q^{\alpha\beta}),\nonumber\\
q^{\alpha\beta}&&=G^{\alpha\beta}(\tau-\tau')\delta_{\mu\nu},
\end{eqnarray}
which is purely static (i.e., ``imaginary-time" independent)
and vanishes on the superconducting--glass--paracoherent
phase boundary. However, for the quantum problem the time dependent
fluctuations of the replica diagonal components
$G^{\alpha\alpha}_{\mu\nu}(\tau-\tau')$ must be considered in
the `` imaginary" Matsubara time $\tau $. More precisely, the replica
diagonal part  $G^{\alpha\alpha}$ is not an order parameter because its
expectation value is non-zero on both sides of the transition; nonetheless it
has to be determined self-consistently along with
the glass  order parameter $q^{\alpha\beta}$.
%%%%%%%%%%%%%%%%%%%%%%%%%%%%%%%%%%%%%%%%
\section{Calculation of the  phase diagram}
%%%%%%%%%%%%%%%%%%%%%%%%%%%%%%%%%%%
A  general    solution
of the self-consistency equation (\ref{matrix})
 poses a rather difficult  problem
since the quantum-mechanical nature of the problem
requires that the time-dependence  of replica-diagonal dynamic parameters
$G^{\alpha\alpha}_{\mu\nu}(\tau-\tau')$ have to be determined
  self-consistently.
Therefore,  we employed here the static {\it ansatz}
(cf. Ref.\onlinecite{bray}) which retains only the $\omega_{\ell=0}$
Fourier component $r\delta_{\mu\nu}=(1/\beta)
\int_0^\beta G^{\alpha\alpha}_{\mu\nu}(\tau)$ of the dynamic
self-interaction. Furthermore, since we are here interested mainly
in the critical line separating the glass and paracoherent
phases (where $q_{EA}=0$) we employ the  replica--symmetry
assumption ($q^{\alpha\beta}\equiv q$ for $\alpha\neq\beta$).
In this way we avoid the subtle intricacies of the replica symmetry breaking
(which, however, will be important {\it inside} the glass phase region).
 Therefore,
\begin{equation}
G^{\alpha\beta}_{\mu\nu}(\tau)=
\left[r\delta_{\alpha\beta}+q(1-\delta_{\alpha\beta})\right]
\delta_{\mu\nu}
\end{equation}
and the effective Hamiltonian becomes
\widetext
\begin{equation}
\hat{H}_{\rm eff}(\tau,\tau')  =
 -\frac{E_J^2}{2}  \sum_{\alpha\beta}\sum_\mu
\left[2r\delta_{\alpha\beta}+2q(1-\delta_{\alpha\beta})\right]S^\alpha
_{\mu}(\tau)S^\beta_{\mu}(\tau').
\end{equation}
\narrowtext
Consequently, the averaged free energy density (\ref{fen}), in the  replica
$n\to 0$ limit, becomes
\begin{eqnarray}
\beta {\cal F}_{\rm av}&& =\frac{1}{2}(\beta E_J)^2 (r^2-q^2)+
\nonumber\\
&&-\int_{0}^{\infty}\sigma d\sigma e^{-{\sigma^2}/{2}}
\ln Z(\sigma),
\nonumber\\
Z(\sigma)&&=\int_{0}^{\infty}\rho d\rho
e^{-{\rho^2}/{2}}
{\cal Z}(\sigma,\rho),
\nonumber\\
{\cal Z}(\sigma,\rho)&&=
{\rm Tr}_{\phi}\exp\left[-\beta {\cal H}_\phi(\sigma,\rho)\right],
\nonumber\\
{\cal H}_\phi(\sigma,\rho)&&=-
\frac{K}{2}\frac{\partial^2}{\partial\phi^2}+\nonumber\\
&&-E_J\sqrt{2}\left(\sigma\sqrt{q}+ \rho\sqrt{r-q}\right)\cos(\phi).
\label{endfree}
\end{eqnarray}
Here we have employed integrations over the auxiliary variables $\sigma,\rho$
and ${\rm Tr}_{\phi}\dots=\sum_\ell \langle\Psi_\ell(\phi)|\dots
|\Psi_\ell(\phi)\rangle$, with $|\Psi_\ell(\phi)\rangle$ being the
eigenstates of the operator ${\cal H}_\phi(\sigma,\rho)$ and
\begin{eqnarray}
q&&=\int_{0}^{\infty}\sigma d\sigma e^{-{\sigma^2}/{2}}
\langle\cos(\phi)\rangle_\sigma^2,
\nonumber\\
r&&=\int_{0}^{\infty}\sigma d\sigma e^{-{\sigma^2}/{2}}
\langle\cos^2(\phi)\rangle_\sigma
\end{eqnarray}
where
\begin{eqnarray}
\langle \dots\rangle_\sigma&&=\frac{1}{Z(\sigma)}\int_{0}^{\infty}\rho d\rho
e^{-{\rho^2}/{2}}\nonumber\\
&&\times
{\rm Tr}_{\phi}
\dots\exp\left[-\beta {\cal H}_\phi(\sigma,\rho)
\right].
\end{eqnarray}

\par
The freezing temperature, i.e. the onset of the
glassy phase, is marked by a non-zero value of the spin glass
order parameter $q$. We can now establish  the equation for the critical line
$T_c(K)$ by expanding the free energy in powers of $q$ and equating the
coefficient of $q^2$ to zero. We find that
\begin{equation}
\beta_c E_J r(\beta_c,{ K})=1,
\label{rc}
\end{equation}
and the self-consistency equation for $r$ is (cf. Eq.(\ref{endfree}))
\begin{equation}
r=\frac{\displaystyle
\int_{0}^{\infty}\rho d\rho e^{-{\rho^2}/{2}}{\rm Tr}_{\phi}
\cos^2(\phi)\exp\left[-\beta {\cal H}_\phi(\rho,0)\right]}
{\displaystyle
\int_{0}^{\infty}\rho d\rho e^{-{\rho^2}/{2}}{\rm Tr}_{\phi}
\exp\left[-\beta {\cal H}_\phi(\rho,0)\right]}
\end{equation}
with the effective single site quantum rotor Hamiltonian
\begin{equation}
{\cal H}_\phi(\rho,0)=- \frac{K}{2}\frac{\partial^2}{\partial\phi^2}-
\mu\cos(\phi),
\label{rotor}
\end{equation}
where $\mu=E_J\rho\sqrt{2r}$.
Finally, using partial integrations
it is convenient to represent the  self-consistency equation for $r$ as
\begin{equation}
\int_{0}^{\infty}\rho d\rho e^{-{\rho^2}/{2}}{\cal Z}(\rho,0)
\left[2(\beta E_J)^2r^2+3-\rho^2\right]={\cal Z}(0,0),
\label{endcrit}
\end{equation}
where ${\cal Z}(0,0)=\theta_3(0,\beta K/2)$,
and $\theta_3(0,\beta K/2)$ is the theta function
\begin{equation}
\theta_3(z,t)=1+2\sum_{k=1}^\infty\cos(2kz)e^{-k^2t}.
\end{equation}
To proceed, we have to  explicitly evaluate ${\cal Z}(\rho,0)$
which amounts to finding  the eigenenergy of the effective quantum
rotor Hamiltonian (\ref{rotor}) as a function of (arbitrary positive) $\rho$.
This is a standard eigenvalue problem involving a Mathieu type
differential equation. Unfortunately, it is difficult
to obtain a useful analytical solution for general values
of $K$ and $\mu$ in this way. Therefore, we treat the
problem in various charging energy-temperature regimes
using a combination of calculational approaches to construct
the entire  phase diagram.
%%%%%%%%%%%%%%%%%%%%%%%%%%%%%%%%%%%%%%%%%%%%%%%%%%%%%%%%%%%%%%%%
\subsection{Semiclassical WKB  limit}
%%%%%%%%%%%%%%%%%%%%%%%%%%%%%%%%%%%%%%%%%%%%%%%%%%%%%%%%%%%%%%%%%%%%%%%
For $k_BT \gg E_J$, $K/E_J \ll 1$ the charging energy is expected
to play a subdominant role. Our approach here is to consider the
``potential energy" $\mu\cos(\phi)$ in the quantum rotor Hamiltonian
(\ref{rotor}) as the leading contribution and treat charging energy
 perturbatively. At  lowest order in $K$ one obtains\cite{jose}
\begin{equation}
{\cal H}_\phi(\rho)\approx -
E_J\rho\sqrt{2r}\left[1-({\beta K}/{24})\right]\cos(\phi).
\label{semix}
\end{equation}
Furthermore, by using the identity
\begin{equation}
I_n(\lambda)=\int_0^{2\pi}\frac{d\phi}{2\pi}e^{\lambda\cos(\phi)}\cos(n\phi),
\end{equation}
where $I_n(\lambda)$ is the modified Bessel function of $n$th-order.
Using Eq.(\ref{endcrit}), the critical boundary for small $K$ is readily
obtained from
\begin{equation}
\int_{0}^{\infty}\rho d\rho e^{-{\rho^2}/{2}}
\left(5-\rho^2\right)I_0\left[
 \rho\sqrt{2\beta E_J}\left(1-\frac{\beta K}{24}\right) \right]=1.
\label{semi}
\end{equation}
For $K=0$ at the classical critical point
we have $k_BT_c/E_J=\sqrt{2}/2\approx 0.71$ (cf. Ref.\onlinecite{remark3}).
%%%%%%%%%%%%%%%%%%%%%%%%%%%%%%%%%%%%%%%%%%%%%%%%%%%
\subsection{Variational method}
%%%%%%%%%%%%%%%%%%%%%%%%%%%%%%%%%%%%%%%%%%%%%%%%%%%%%
The quantum mechanical partition function ${\cal Z}(\rho,0)$
in Eq. (\ref{endcrit}) can be approximated by an effective
classical function via the path integral formalism using
a variational method. Following Refs.\onlinecite{Kleinert,Vaia}
(see Appendix for details pertaining this problem)
we obtain the critical  boundary
between the glassy and paracoherent phases from
\widetext
\begin{equation}
\int_0^\infty d\rho \rho e^{-\rho^2/2+\xi(\rho)}
\frac{\beta K\Omega_c(\rho)(5-\rho^2)/2}{\sinh\left[\beta K\Omega_c(\rho)/2
\right]}I_0\left[\rho\sqrt{2\beta E_J}
\exp\left(-\frac{a_c^2(\rho)}{2}\right)\right]=1,
\label{vario}
\end{equation}
\narrowtext
where $\xi_c$, $a^2_c$ and  $\Omega^2_c$ are determined (for a given
value of $\rho$) from the set of self-consistency equations:
\begin{eqnarray}
\xi_c(\rho)&=&\frac{\beta_c K}{2}\Omega^2_ca^2_c,
\nonumber\\
a^2_c&=&\frac{1}{\beta K\Omega_c^2}\left[\frac{\beta K\Omega_c}{2}
\coth\left(\frac{\beta K\Omega_c}{2}\right)-1\right],
\nonumber\\
\Omega^2_c&=&\rho\frac{\sqrt{2\beta_c E_J}}
{\beta_c K}\exp\left(-\frac{a^2_c}{2}\right).
\label{yyy}
\end{eqnarray}
Numerical evaluation of Eq.(\ref{vario}) reveals a
reentrance in the low temperature region from the glassy phase
back to the paracoherent state (see Fig.2)
and the zero-temperature critical value
of reduced charging energy $\alpha_c=K/E_J\approx 1.1$
In the opposite semiclassical limit, $K\to 0$
Eq.(\ref{vario}) reduces to Eq.(\ref{semi}) (see Appendix).
The important question arises as to whether the
predicted reentrant feature is a genuine property
of the model or it is  an artifact of the approximation.
We note that the variational method
implementation\cite{Kleinert,Vaia} considers that
the paths must satisfy the boundary condition
$\phi(0)=\phi(\beta)$, appropriate  for
a quantum particle in a periodic potential  rather than  a
quantum rotor system (Eq.(\ref{rotor})) with $2\pi$ periodic wave functions.
Therefore, the  proper boundary conditions for the
latter  must be $\phi(0)=\phi(\beta)+2\pi m$ (where
$m=0,\pm 1,\pm 2,\dots$ are the winding numbers)
which appear not to  be accounted for consistently in this approach.
This might have consequences for the low temperature
behavior of the system and we are therefore  motivated
to look for  an alternative method to study the quantum
$T\to 0$ limit.

As we shall see next, the reentrant transition found in the
variational approximation is not found in the low temperature expansion.
Thus it is most likely an artifact of the variational approach,
which is strictly a high temperature approximation.

%%%%%%%%%%%%%%%%%%%%%%%%%%%%%%%%%%%%%%%%%%%%%%%%%%%%%%
\subsection{Perturbation expansion about the quantum  limit}
%%%%%%%%%%%%%%%%%%%%%%%%%%%%%%%%%%%%%%%%%%%%%%%%%%%%%%
Assuming that$K \gg E_J$ the potential energy of the quantum rotor
(\ref{rotor}) $\mu\cos(\phi)$ may be treated perturbatively
using the standard Rayleigh-Schr\"{o}dinger approach.
The unperturbed part is then $-(K/2)\partial^2/\partial\phi^2$
with eigenfunctions
\begin{equation}
|\Phi^0_m(\phi)\rangle=\frac{1}{\sqrt{2\pi}}\exp(im\phi).
\label{wave}
\end{equation}
To lowest nontrivial order we get
\begin{equation}
E_m(\rho)=\frac{K}{2}\left[m^2-\frac{4\left(\displaystyle
\frac{E_J}{K}\right)^2r\rho^2}{1-4m^2}\right],
\end{equation}
so that
\begin{equation}
{\cal Z}(\rho )=\sum_{m=-\infty}^{m=+\infty}e^{-\beta E_m(\rho)},
\end{equation}
and the critical boundary condition
is implemented after performing the integration obtaining
\widetext
\begin{equation}
\sum_{m=-\infty}^{m=+\infty}e^{-m^2\beta_c K/2}
\frac{16m^4+8m^2\left[3(E_J/K)-1\right]-8(E_J/K)^2-6(E_J/K)+1}
{\left[4m^2+4(E_J/K)-1\right]^2}=0.
\end{equation}
\narrowtext
At $T=0$ the $m=0$ term is the only one that  contributes and thus
\begin{equation}
\frac{1}{1-4\left({E_J}/{K_c(T=0)}\right)}
\left[5-\frac{2}{1-4\left({E_J}/{K_c(T=0)}\right)}\right]=1,
\end{equation}
that  gives $\alpha_c(T=0)=(6+2\sqrt{17})/2\approx 7.1231$
which differs quite significantly from the critical value of
$\alpha$ obtained previously by the variational method\cite{wires1,wires2}.
We attribute this difference again to the fact that in
the  present formulation of the variational method\cite{Kleinert,Vaia}
the boundary conditions for the superconducting phase
variables are not compatible with the  discrete nature of the
charge transfer process which becomes especially important as $T\to 0$.
As a result, a discrepancy in the low temperature limit is expected, whereas
in the classical limit ($K\to 0$) $T_c$ is reproduced correctly.
%%%%%%%%%%%%%%%%%%%%%%%%%%%%%%%%%%%%%%%%%%%%
\subsection{Truncated charge space projection method}
%%%%%%%%%%%%%%%%%%%%%%%%%%%%%%%%%%%%%%%%%%%%%
The expansion about the quantum limit from the previous subsection
is not sufficient to treat the critical boundary away from
the quantum critical point $\alpha_c(T=0)$, where
the condition $\alpha\gg 1$ is no longer valid.
A non-perturbative approach is  sought and
of  particular interest are the truncated charge
state models (TSCM) spanned by the charge states
of the operator $-(K/2)\partial^2/\partial\phi^2$
in a restricted, finite-dimensional,  Hilbert space.
This can be interpreted as an approximation
of a suitable  quasi-spin model.
Using the eigenstates of the charge operator (\ref{wave})
one can readily calculate the matrix elements of the
operators involved
\widetext
\begin{eqnarray}
N_{km}&&=\langle\Phi_k(\phi)|\hat{n}(\phi)|\Phi_m(\phi)\rangle=
\int^{2\pi}_0\frac{d\phi}{2\pi}\exp(-ik\phi)
\left(\frac{2e}{i}\frac{\partial}{\partial\phi}\right)
\exp(im\phi)=2e\, m\delta_{km},\nonumber\\
{[S_x]}_{km}&&=\langle\Phi_k(\phi)|\cos(\phi)|\Phi_l(\phi)\rangle=
\int^{2\pi}_0\frac{d\phi}{2\pi}\exp[-i(k-m)\phi]\cos(\phi)\nonumber\\
&&=\frac{1}{2}(\delta_{k-m-1,0}+\delta_{k-m+1,0}).
\end{eqnarray}
\narrowtext
In particular, for $k,m=0,\pm1$, one has
\begin{eqnarray}
N_{km}&&=[{\cal S}_z]_{km},\nonumber\\
{[S_x]}_{kl}&&={\left[\frac{{\cal S}_x}{\sqrt{2}}\right]}_{km},
\end{eqnarray}
where ${\cal S}_a (a=z,x)$ are given by
\begin{equation}
{\cal S}_z=
\left(
\begin{array}{ccc}
1 & 0 & 0\\
0 & 0 & 0 \\
0 & 0 & -1
\end{array}
\right),\quad\quad
{\cal S}_x=
\frac{1}{\sqrt{2}}
\left(
\begin{array}{ccc}
0 & 1 & 0\\
1 & 0 & 1 \\
0 & 1 & 0
\end{array}
\right).
\end{equation}
Consequently, the lowest-order quasi-spin model belongs to ${\cal S}=1$
and its Hilbert space is spanned by the charge states
$ |0\rangle,|\pm 1\rangle$. It is now useful to
recast the quantum rotor Hamiltonian (\ref{rotor})
in the ${\cal S}=1$  pseudospin language as
\begin{equation}
H=(K/2){\cal S}_z^2-(\mu/\sqrt{2}){\cal S}_x.
\label{spinham}
\end{equation}
In the context of magnetic spin glasses the  first charging energy term
in Eq.(\ref{spinham}) refers to a single-ion crystal
anisotropy, which opposes  ordering in the $x-y$
plane\cite{kopec2}. Thus the system will exhibit a phase transition
driven by quantum fluctuations of the transverse (pseudo)spin component.
Correspondingly, for the statistical sum ${\cal Z}(\rho,0)$ we have
\begin{eqnarray}
&&{\cal Z}(\rho,0)\equiv
{\cal Z}_{S=1}(\rho,0)=\exp\left(-{{{\beta_c K}\over 2}}\right)+
\nonumber\\
&&+2\exp\left(-{{{\beta_c K}\over 4}}\right)
\cosh\left(\frac{\sqrt{(\beta_c K)^2+ (\beta_c E_J)^2\rho^2 }}{4}\right),
\label{lala}
\end{eqnarray}
while
\begin{equation}
\int_{0}^{\infty}\rho d\rho e^{-{\rho^2}/{2}}{\cal Z}(\rho,0)
\left(5-\rho^2\right)={\cal Z}(0,0),
\label{endcrit2}
\end{equation}
and the critical line $T_c(\alpha)$ can be  readily calculated
numerically using Eq.(\ref{endcrit2}) (see Fig.\ref{fig2}).
We found $\alpha_c(T=0)=7$ in good agreement with the perturbative
approach. The classical value $T_c(\alpha=0)$ is underestimated
as a result of restricting the original charge state
Hilbert space. We have also examined
 models  spanned by five  $(|0\rangle,|\pm 1\rangle,
|\pm 2\rangle)$ and seven
$(|0\rangle,|\pm 1\rangle,|\pm 2\rangle,|\pm 3\rangle)$
charge states
which better approximate the behavior at high temperatures.
For both cases it is possible to derive critical line
equations analogous to Eq.(\ref{lala}) analytically. However,
the corresponding formulae are too lengthy to  reproduce here.
We found that
close to $T_c(\alpha=0)$ TCSM exhibit a small reentrance
for certain interval of $\alpha$'s. However, as the number
of charge states $m_{\rm charge}$ increases this interval became narrower
and presumably disappears for $m_{\rm charge}\to\infty$
indicating that
this might be a spurious feature due to  the restriction
imposed on the original infinite-dimensional Hilbert space
of charge states.

%%%%%%%%%%%%%%%%%%%%%%%%%%%%%%%%%%%%%%%%%%%%%%%%%%%
\section{Conclusions}
%%%%%%%%%%%%%%%%%%%%%%%%%%%%%%%%%%%%%%%%%%%%%%%%%%%

In this paper we have studied the competition
between quantum and thermal fluctuations
in a superconducting glass state
using   positionally disordered model of
ulthrathin Josephson coupled wires in a transverse
magnetic field.
We focused  our attention in the regime where
the magnetic field produces large frustration
in a system of  randomly  spaced
Josephson microjunctions with large number of nearest-neighbors
$z=N$. We investigated this model by spin-glass inspired techniques
in the $N\to\infty$ limit, showing that  that the model  is a
superconducting analogue of a quantum spin-glass with
Dzialoshinsky-Moriya time-reversal braking interaction.
\par
We studied  the phase boundary
separating the paracoherent from glassy phases as a function
of the charging energy associated with the  small capacitance
of an individual wire. We found that, for sufficiently
large charging energy, the glassy ground state
is  destabilized due to the strong quantum
fluctuations. This is reminiscent of the scenario
found in ordered Josephson junction arrays, where
charging energy effects can lead to the  destruction
of long-range phase coherence by zero-point quantum fluctuations.
However, this analogy is not complete as in the glassy phase there is
no long-range order  and, therefore,  the destructive role of quantum
fluctuations is less transparent. We note that it is not easy to improve
our static {\it ansatz} analytically. We expect that a direct
Quantum Monte Carlo calculation, where the imaginary time direction is
discretized, will yield information inside the phase boundary. The
shortcoming of this approach is that one can not get too close to the
$T=0$ region. We leave this problem for the future.
\par
 The results presented here can be tested
experimentally in artificially fabricated disordered arrays of ultrathin
filaments provided that the charging energy of a wire
is large enough to produce substantial superconducting quantum
phase fluctuations.
For example, $T_c(\alpha)$ should be observable
by resistivity measurements. In obtaining our results for the phase
diagram we considered the solution which does not break the replica symmetry.
Although a broken replica symmetry solution is not required to trace the
critical phase boundary (where the order parameter vanishes)
it is  important when distinguishing  equilibrium from
nonequilibrium properties of the glassy phase. In the
low-temperature phase, as usual, the
strongest signatures of  glassiness are in the dynamical
properties. Similar to the magnetic spin glasses,
non-ergodic behavior is likely to manifest itself
in differences between field-cooled and zero-field
prepared samples via typical effects like hysteresis,
remanence, aging effects etc.
In the classical limit
of our model (vanishing of charging energy) all these
history dependent signatures of a glassy state have
been observed\cite{wires2,choi2,joseglass}.
The corresponding non-equilibrium metastable states might be probed
eg. via {\it ac} conductivity measurements  determining the
barriers separating metastable states and the
associated distribution of relaxation times as a function
of charging energy.
\par
Although the system of superconducting
ultrathin wires constitutes an interesting experimental setting
to test  the predictions of the infinite-range
interaction theory, the investigation
of the behavior of  disordered quantum
truly  $2D$ array  would require
knowledge relevant to short-range spin glasses and
presents a difficult subject for further study.

%%%%%%%%%%%%%%%%%%%%%%%%%%%%%%%%%%%%%%%
\acknowledgments
This work has been partially supported  by  NSF grants DMR-95-21845
and PHYS-94-07194 (ITP at UC, Santa Barbara),
and  by  the US Information Agency (Senior Fulbright Scholarship, TKK).
%%%%%%%%%%%%%%%%%%%%%%%%%%%%%%%%%%%%
\appendix
%%%%%%%%%%%%%%%%%%%%%%%%%%%%%%%%%%%%%%%%%%%%%%%%%%%
\section{Variational method}
%%%%%%%%%%%%%%%%%%%%%%%%%%%%%%%%%%%%%%%%%%%%%%%%%%%%%
In this appendix we give the specific details of the derivation
of Eq. (39) based on the variational approach\cite{Kleinert,Vaia}.
The path integral formulation of the partition function
involves an infinite product of periodic paths
$\phi(\tau)=\phi_0+\sum_{\ell=1}^\infty(\phi_\ell
 e^{i\omega_\ell\tau}+c.c)$
in the form
\widetext
\begin{eqnarray}
{\cal Z}(\rho,0)&&=\int
\frac{d\phi_0}{\sqrt{2\pi\beta K}}
\prod_{\ell=0}^{\infty}\left[\int
\frac{d\phi^{\rm re}_\ell d\phi^{\rm im}_\ell}
{\pi/( \beta K\omega^2_\ell)}\right]\times\nonumber\\
&&\times
\exp\left\{
-\beta{K}\sum_{\ell=1}^{\infty}\omega^2_\ell|\phi_\ell|^2
-{\mu}\int_0^{\beta}d\tau\cos\left[\phi_0+\sum_{\ell=1}^{\infty}
(\phi_\ell e^{-i\omega_\ell}+ {\rm c.c})\right]\right\}.
\label{statsum}
\end{eqnarray}
\narrowtext
The essence of the variational method (\cite{Kleinert,Vaia})
is to approximate (\ref{statsum})
by an  effective {\it classical} partition function
\begin{equation}
{\cal Z}(\rho,0)=\int\frac{d\phi_0}{\sqrt{2\pi\beta K}}
e^{-\beta V_{\rm eff}(\phi_0)}.
\end{equation}
Here, the associated effective potential $V_{\rm eff}$ is given by
\begin{equation}
V_{\rm eff}=\frac{1}{\beta}\ln\left[\frac{\sinh(\beta K\Omega/2)}
{\beta K\Omega/2}\right]
 +V_{a^2}-\frac{K}{2}\Omega^2 a^2.
\end{equation}
The unknown functions $a^2(\phi_0)$, $V_{a^2}(\phi_0)$ and $\Omega(\phi_0)$
are determined by using extremal principle from the self-consistency
conditions
\begin{eqnarray}
a^2&&=\frac{1}{\beta K\Omega^2}\left[\frac{\beta K\Omega}{2}
\coth\left(\frac{\beta K\Omega}{2}\right)-1\right],
\nonumber\\
V_{a^2}&&=-\mu\int^{+\infty}_{-\infty}\frac{d\phi}{\sqrt{2\pi a^2}}
\exp\left[-\frac{(\phi-\phi_0)^2}{2a^2(\phi_0)}\right]\cos(\phi),
\nonumber\\
&&= -\mu\exp\left(-\frac{a^2}{2}\right)\cos(\phi_0),
\nonumber\\
\Omega^2&&=\frac{\mu}{K}\exp\left(-\frac{a^2}{2}\right)\cos(\phi_0).
\label{xxx}
\end{eqnarray}
For low temperatures we seek for the uniform solution
 (i.e., with $\phi_0=0$) of Eq.(\ref{xxx}) (cf. Eq.(\ref{yyy})). In the regime
of  high temperature and small quantum fluctuations by using the expansion
\begin{equation}
\frac{1}{x}\left[\frac{x}{2}\coth\left(\frac{x}{2}\right)-1\right]
=\frac{x}{12}-\frac{x^3}{720} +O(x^4),
\end{equation}
we have for $V_{\rm eff}(\phi_0)$
\begin{eqnarray}
{\cal H}(\rho,0)&&\rightarrow -E_J\rho\sqrt{2r}\exp\left(-\frac{\beta K}{24}
\right)\cos(\phi_0),\nonumber\\
&&\approx -E_J\rho\sqrt{2r}\left(1-\frac{\beta K}{24}
\right)\cos(\phi_0),
\end{eqnarray}
i.e., $V_{\rm eff}(\phi_0)$ reduces to semiclassical WKB result (\ref{semix}).
%%%%%%%%%%%%%%%%%%%%%%%%%%%%%%%%%%%%%%%%%%%%%%%%%%%%%%%

\begin{figure}
\caption{Disordered Josephson-coupled
array of $2N$ superconducting  wires (straight horizontal
and vertical  lines, square box denotes the junction at a node).
Each wire has small self-capacitance $C$
and is characterized by the superconducting phase
$\phi_{i}$. The magnetic field $B$ is applied perpendicular
to the array.}
\label{fig1}
\end{figure}
\begin{figure}
\caption{Temperature-charging energy, $T-\alpha$,
phase diagram of the
disordered Josephson-coupled array of superconducting wires with
self-charging energies.  Here $\alpha =\frac{K}{E_J}$,
with  $K$  the charging energy and $E_J$
the Josephson coupling energy. The results were
obtained from variational method (VM)
and truncated charge states models (TCSM) for different numbers
of charge states as indicated. We note, as discussed in the text, that
the reentrant behavior obtained from the variational calculation does
not emerge from the low temperature analysis and thus it must be an
artifact of the variational approximation.}
\label{fig2}
\end{figure}

\begin{references}
\bibitem{remark2}
in the recent work (see Ref. \onlinecite{recent3})
 on classical Sherrington-Kirkpatrick
superconducting glass it was assumed differently that $f_{ij}$ are
Gaussian distributed: $P(f_{ij})\sim exp(-f_{ij}^2/2\sigma)$
which reduces to the case of uniform distribution of $f_{ij}$
considered here only in the limit  $\sigma\to\infty$.
\bibitem{remark3}
Note that the classical critical temperature
found here is higher then the value found in
Ref.\onlinecite{recent} $(T/J=0.5)$ for the
disordered array. Similar
discrepancy has been reported  in the Ref.
\onlinecite{wires1} where Monte Carlo
simulations gave  $(T/J=0.75)$. We believe
that this difference is due to an simplified assumption
made in Ref.\onlinecite{recent}  that the
gauge--glass  can be mapped onto a pure
$XY$ spin--glass problem (without the time--reversal breaking
Dzialoshinsky--Moriya interaction,
see  Eq.\ref{dm}  and the  argumentation below).
\end{references}
\end{document}